\documentclass[%
 reprint,
 amsmath,amssymb,
 aps,
superscriptaddress
]{revtex4-2}
\usepackage{amsthm}
\usepackage{color}

\newtheorem{theorem}[]{Theorem}

\usepackage{graphicx}
\usepackage{dcolumn}
\usepackage{bm}
\definecolor{pdv}{rgb}{1,0,0}

\begin{document}

\preprint{APS/123-QED}

\title{Strong Gravitational Lensing in Horndeski theory of Gravity}

\author{Pedro Bessa}
 \email{pedvbessa@gmail.com} 
\affiliation{PPGCosmo, CCE - Federal University of Esp\'\i rito Santo, zip 29075-910, Vit\'oria, ES, Brazil.}

\affiliation{Department of Theoretical Physics, Universit\'e de Gen\`eve, Quai
E. Ansermet 24 , Gen\`eve, 1211, Switzerland.}
 \email{pedvbessa@gmail.com}

\date{\today}

\begin{abstract}
In this paper we build the general gravitational lensing formalism for luminal Horndeski theories, deriving the Jacobi matrix equation and the general angular diameter distance in these theories through the screen space formalism. We generalize the focusing and multiple-lensing theorems in General Relativity to include the luminal Horndeski theories and derive constraints they must satisfy to exhibit the same gravitional lensing behavior in General Relativity. This provides a way to test theories through Strong Lensing effects, as well as a full theoretical framework for testing lensing predictions in these theories against observations. We find that for some theories, like metric $f(R)$ and unified $k$-essence, the obtained theorems are satisified in general physical cases, while for others, like Galileon Condensate models, the current observational constraints show the theory has the same gravitational lensing behavior as in General Relativity.

\end{abstract}

\maketitle


\section{Introduction}

Gravitational Lensing promises to be a powerful probe of Gravitation on large scales, with weak lensing by clusters and large scale structure providing tests of the concordance cosmological model \cite{Mandelbaum_2018,Umetsu_2020} and strong lensing by Black Holes and compact objects providing tests of Gravity on small scales beyond Solar System constraints \cite{EHT_19,ethCollaboration_2022}.

The search for a solution to the nature of Dark Energy has led to intense research in Scalar-Tensor theories and their behavior in the cosmological setting \cite{Clifton_2012}. Since these theories in general modify the gravitational coupling and energy content of gravity, one would expect deviations from the behavior predicted by General Relativity. Beyond the usual PPN formalisms \cite{Keeton2005}, the deviation from GR  should be derived from principle, starting from the Modified Theory.

Developing a rigorous approach to the behaviour of gravitational lensing in Modified Gravity is important when new lensing regimes become accessible through advances in observational capabilities, with both the current and next generation of surveys expected to increase the statistics of strong gravitational lensing in a $10^5$-fold way \cite{Jacobs_2019}. Ever growing precision in observations requires a full theory to distinguish the pure relativistic effects arising from GR from the possible effects of modifications of gravity.

The study of imprints of Modified Gravity in Gravitational lensing dates back to Bekenstein \cite{Bekenstein_1994}, which predicted the expected light bending for nonminimally coupled theories and their underestimation of the mass in galaxy clusters. Research on TeVeS and MOND-like theories and their effects on both weak and strong gravitational lensing has been extensive \cite{Mortlock_01,moffat2009,Moffat_2021,Moffat_2018,Rahvar_2018}, while theories of the Jordan-Brans-Dicke type have been explored in \cite{Keeton2005,Keeton2006} using the PPN formalism; in \cite{Schimd_2005,Gao_2019} for spacetimes in the weak field limit and perturbed cosmologies; and in \cite{Campigotto_2017,Alhamzawi_16} in general spherically symmetric spacetimes for specific theories. More recently, there's been interest in gravitational lensing in general Scalar Tensor theories for black hole solutions and supermassive black holes (SMBH), such as in \cite{Chagoya_2021,Kumar_2022}, and for charged black holes in \cite{Wang_2019}. Observational tests and constraints of modified gravity through weak lensing, mainly using parametrized perturbations, can be found in \cite{Schmidt_2008,Tereno_11}, and recently, using the EHT observations, in \cite{Afrin_2022}.

While these studies deal with specific theories and regimes, there's been a lack of a systematic and rigorous treatment of lensing in general Modified Gravity theories. The present paper attempts to fill that gap, developing the mathematical formalism necessary to deal with Gravitational Lensing in the class of Luminal Horndeski theories, the most general 2nd order Scalar Tensor theories with non-degenerate Lagrangian and luminal tensor propagation speed, which include theories such as quintessence, $f(R)$, Brans-Dicke, k-essence and cubic galileons \cite{Kobayashi_2019,Kase_2013}. 

We develop our formalism from the top down, first describing the general behavior of light bundles in modified gravity theories using an effective geometrical stress-energy tensor $T_{\mu\nu}^{\text{eff}}$. We then derive the Jacobi equation and its immediate consequences, the focusing and lensing equations, which dictate the behavior of light rays in the general lensing regime \cite{Korzynski_2018}, their stretching, magnification and deflection. We then prove a couple of theorems that extend the focusing and multiple image theorems for General Relativity, under general weak energy and average energy condition assumptions \cite{hawking_ellis_1973}. Finally, w1e discuss how the detection of lensing effects that depart from the General Relativity predictions can be used as constraints on the parameter space of certain theories.

The paper is structured as follows: In section \ref{Horndeski_EFE}, we review the Horndeski theory of Gravity, its field equations and luminal limit; in \ref{lensing_GR} we review the basic mathematical formalism of gravitational lensing in General Relativity; in section \ref{lensing_horndeski} we adapt this formalism to Horndeski theories and obtain the focusing and lensing equation in arbitrary spacetimes. We also obtain the main theorems of the paper and test their assumptions against 4 classes of theories in the Horndeski family. Finally, in \ref{conclusions} we discuss possible uses of the formalism and how the results can put constraints in Horndeski theories and test Modified Gravity using lensing.

\section{Horndeski Gravity and Field Equations}
\label{Horndeski_EFE}
In \cite{Horndeski1974} the most general stable Scalar-Tensor lagrangian with second order equations of motion was obtained. In \cite{Kobayashi_2011}, this Lagrangian was rediscovered in the context of Inflation and in connection to the so called Generalized Galileon theories \cite{De_Felice_2011}. The generality and stability of the theory provided the basis for the Effective Field Theory of Dark Energy \cite{Gubitosi_2013,Frusciante_2020} and other effective approaches, which have been developed as a standard way to treat deviations from GR in the cosmological setting \cite{Bellini_2014}.

In this work, we use the Lagrangian formulation of the theory using the so called Horndeski functions. The other approaches, such as the Effective Field Theory of Dark Energy (EFTDE), while useful in certain settings, are not suited for the generality that we require in this paper; for instance, these approaches often require that the space-time has a well defined ADM decomposition \cite{Frusciante_2020}. 
Using the convention of \cite{Bellini_2014}, the Horndeski Lagrangian can be written in the form 

\begin{eqnarray}
    S & = & \int d^4x \sqrt{-g} \left(\sum_{n=1}^{5}\mathcal{L}^{(n)} \right),\\
    \mathcal{L}^{(1)} & = & \frac{1}{2}R, \quad \mathcal{L}^{(2)}  = G_2\left(X,\phi\right), \quad \mathcal{L}^{(3)} = -G_3\left(X,\phi\right)\Box\phi,\\
    \mathcal{L}^{(4)} & = & G_{4}\left(X,\phi\right)R+ G_{4X}\left(X,\phi\right)\left[\left(\Box\phi\right)^{2}-\left(\nabla_{\mu}\nabla_{\nu}\phi\right)^{2}\right]\\
    \mathcal{L}^{(5)} & = &G_{5}\left(X,\phi\right)G_{ab}\nabla^{a}\nabla^{b}\phi\\& - &\frac{G_{5X}\left(X,\phi\right)}{6}\left[\left(\Box\phi\right)^{3}-3\Box\phi\left(\nabla_{a}\nabla_{b}\phi\right)^{2}
    +2\left(\nabla_{a}\nabla_{b}\phi\right)^{3}\right], \notag
\end{eqnarray}
where we have explicitly separated the pure GR density $R/2$ from the Horndeski density $\mathcal{L}^{(4)}$, against convention. This will be useful when defining effective tensors. We define  $X\equiv -\nabla_\mu \phi \nabla^\mu \phi/2$, and $G_{iX} = \partial_X G_i$.

One also has, in general, the matter field lagrangian, which is coupled only to gravity through the metric

\begin{equation}
    \mathcal{L}^{(m)} = \mathcal{L}(g_{\mu\nu},\Psi),
\end{equation}
$\Psi$ the matter fields of, e.g., perfect fluids, the standard model or radiation.

The $G_5$ and $G_4$ are related to the propagation of gravitational waves \cite{Bellini_2014}, and the recent detection of the gravitational event GW170817 and its electromagnetic counterpart has put tight constraints on the deviation of the propagation speed of gravitational waves from the speed of light \cite{Ezquiaga_2017,Creminelli_2017}. \cite{Kase_2019} and \cite{Ezquiaga_2017} argue that the most natural way to avoid fine-tuning while still demanding that the theories have luminal speed of gravitational waves is to set 

$$G_{4X} = G_{5X} = G_{5\phi} = 0,  $$

which means no kinetic coupling to the curvature, and no tuning in the coupling of the Einstein tensor. From these constraints, the most general Horndeski Lagrangian with propagation speed of tensor modes $c_T = c$ is the one given by the Lagrangian

\begin{align}
\label{Horndeski_luminal}
    \mathcal{L} = \frac{R}{2} +&G_2(X,\phi) -G_3(X,\phi)\Box\phi \\
    +&G_4(\phi)R+G_5 G_{\alpha\beta}\nabla^\alpha\nabla^\beta\phi.\notag
\end{align}
The $G_5$ coupling term is a total derivative, so it can be discarded in the variational derivation of the equations of motion.
In the rest of this paper, when referring to "Horndeski theories" we mean the ones described by the lagrangian \eqref{Horndeski_luminal}

\subsection{Field Equations}
The dynamics of the fields $\phi$ and $g_{\mu\nu}$ are obtained by variation of \eqref{Horndeski_luminal}. We first write them out explicitly, and then separate the parts related to each coupling term in effective stress energy tensors $T_i^{\text{eff}}$

\begin{flalign}
    \label{field_equations_1}
    G_{\mu\nu} =&\notag \\
    &G_2g_{\mu\nu} + G_{2X}\nabla_\mu\phi\nabla_\nu\phi &&\notag\\
    +&G_{3X}\left(\nabla^\alpha\phi\nabla_\alpha Xg_{\mu\nu} -\Box\phi\nabla_\mu\phi\nabla_\nu\phi-2\nabla_{(\mu}\phi\nabla_{\nu)}X\right) \notag\\
    -2&G_{3\phi}\left(X g_{\mu\nu} +\nabla_\mu\nabla_\nu\phi \right)\notag\\
    -2&G_4G_{\mu\nu}+2G_{4\phi}\left(-\Box\phi g_{\mu\nu} + \nabla_\mu\nabla_\nu\phi\right)\notag\\
    +2&G_{4\phi\phi}\left(2X g_{\mu\nu} +\nabla_\mu\phi\nabla_\nu\phi \right).
\end{flalign}

We define the right hand side of equation \eqref{field_equations_1} as a sum of effective stress energy tensors $T_{\mu\nu}^{(i)}$, defined by variation of each term in \eqref{Horndeski_luminal} containing the coupling $G_i$ in terms of the metric:

\begin{equation}
    T_{\mu\nu}^{(i)} \equiv \frac{-2}{\sqrt{-g}}\frac{\delta(\sqrt{-g} \mathcal{L}^{(i)})}{\delta g_{\mu\nu}}.
\end{equation}

Equation \eqref{field_equations_1} is then written as

\begin{equation}
    \label{field_equations_2}
    G_{\mu\nu} = (1+2G_4)^{-1}\left[T_{\mu\nu}^{(2)}+ T_{\mu\nu}^{(3)} + T_{\mu\nu}^{(4)} + T_{\mu\nu}^{(5)}  \right] + T_{\mu\nu}^{\text{(m)}},
\end{equation}
where the last term is the stress-energy tensor of ordinary matter, uncoupled to the scalar field.

Writing the field equations in the form \eqref{field_equations_2} allows us to separate the curvature dependency of the Horndeski interactions to the left side of the equation,  such that the modifications of the geodesic deviation equation (GDE) are written in a straightforward way. In the following section we give a brief review of the mathematics of gravitational lensing in General Relativity to then derive the modified equations for gravitational lensing.

\section{Lensing Formalism in General Relativity}
\label{lensing_GR}
The lensing formalism for arbitrary spacetimes in the case of General Relativity has been thoroughly studied, with classic texts such as \cite{Schneider}, and modern reviews and treatments \cite{Perlick,Fleury_2021,Schneider2006}. In this section, we'll briefly review the basic tools of gravitational lensing formalism in General Relativity in order to extend it to the Horndeski theories.

\subsection{Jacobi map and null geodesics}

For a given geodesic $\gamma$ defined on a spacetime $(\mathcal{M}, g_{\mu\nu})$, with affine parameter $s$ and tangent vector field $\mathbf{k} \equiv \nabla_s \gamma$, we define its geodesic neighbourhood, parameterized by an infinitesimal vector $\xi^\mu$ and a parameter $\epsilon$, as being the set of curves $x(s,\epsilon)$ satisfying

\begin{equation}
\label{bundle}
    \nabla_s x(0,0) = \mathbf{k}, \quad \nabla_\xi x(0,\epsilon) = \epsilon, \quad x(s,0) = \gamma(s). 
\end{equation}

This implicitly defines a map $x:\mathbb{R}^2 \mapsto \mathcal{M}$, its image called the screen space $\mathbf{S}$ \cite{Korzynski_2018}. The deviation vector $\xi^\mu$ is parallely transported through the geodesic bundle, and satisfies the relation

$$\mathcal{L}_\mathbf{k} \mathbf{\xi} =  [k^\mu, \xi^\nu] = 0.$$

From the above relations, one can obtain the Geodesic Deviation Equation

\begin{equation}
\label{GDE}
\frac{D^2 }{d s^2}\xi^\mu = R^\mu_{\alpha \beta \nu} k^\alpha k^\beta \xi^\nu.
\end{equation}

We now define a frame basis for the screen space, which is commonly called the Sachs Basis \cite{Perlick}, satisfying

\begin{equation}
\label{sachs_basis}
    E^A_\mu \in \mathbf{S}, \quad E^A_\mu E^{\mu B} = \delta^{AB}, \quad k^\mu E_\mu^A = 0.
\end{equation}

It is clear that this basis is orthonormal and tangent to the geodesic bundle defined by \eqref{bundle}. The indexes $A\in {1,2}$ label the 2 real dimensions of the parametrization, while the greek indexes label the coordinates in spacetime.
This basis is the one which we measure distortion by the gravitational lenses, providing unitary vectors to which one can measure the lensing angles.

In relation to the basis \eqref{sachs_basis}, we write a vector $y^\mu_A$ in the screen space $\mathbf{S}$ as 

\begin{equation}
\label{jacobi_def}
    y^\mu_A = D_A^B E_B + Y_A k^\mu  .
\end{equation}

Rewriting the vector $\xi^\mu$ in \eqref{GDE} in the Sachs basis and using \eqref{GDE} , we obtain that the matrix $D_A^B$ satisfies the Jacobi matrix equation:

\begin{equation}
\label{jacobi_eq}
    \nabla_k\nabla_k D^A_B = R^B_{\alpha \beta C} k^\alpha k^\beta D^C_A,
\end{equation}
where $\nabla_k = k^\nu \nabla_\nu$.

This equation describes the evolution of the Jacobi matrix on the manifold. Setting initial conditions at the source plane $\mathbf{S}_S$, this defines the mapping of the separation angle $\theta$ of two points, or objects, at the source plane, to the observed angle $\beta$ at the observer plane $\mathbf{S}_O$. We can omit the screen space indices $A,B$ and use the subscript notation $D_{SO}$ to denote a Jacobi matrix that maps a vector in  $\mathbf{S}_S$ to a vector in $\mathbf{S}_O$. It can be shown that $D_{SO} = -D^\text{T}_{OS}$, that is, the Jacobi matrix is anti-hermitean, and therefore diagonalizable with orthogonal eigenvectors.

For a given observer $O$ with 4-velocity $u^\mu$, We define the measured energy of a null ray in the bundle as

\begin{equation}
    E_O= -k^\mu u_\mu|_O, 
\end{equation}

and the redshift $z$ as the ratio

\begin{equation}
  1+z_S \equiv E_S/E_O
\end{equation}

between the energy measured at the event $S$ and the observer $O$ in the worldline of the null ray.

We now consider a thin lens, meaning a space-like hypersurface which is pierced by the null ray bundle of geodesics, defined in \eqref{bundle}, at the lens plane $\mathbf{S}_L$. If two rays separated by an angle $\theta$ at the source plane $\mathbf{S}_S$ cross the lens plane $\mathbf{S}_L$ and are deflected by an angle $\alpha$ then the Lens map, which maps the separation $\theta$ at the source plane to the observed separation $\beta$ at the observer, is given by \cite{Fleury_2021}

\begin{equation}
    \label{lens_map}
    \beta(\theta) = \theta - (1+z_L) D_{OS}^{-1}\left[D_{LS}\left[\alpha\right]\right] (D_{OL}\left[\theta\right]),
\end{equation}

where we note that the $D$ are matrices on the respective vector spaces that span  $\theta$ and $\alpha$.
The deflection angle $\alpha$ is defined in terms of the surface mass density of the lens $\Sigma$, which gives the mass profile of the lens at the lens plane for the thin lens approximation. 
\section{ Gravitational Lensing in Horndeski Gravity}
\label{lensing_horndeski}

\subsection{Strong Gravitational Lensing}

In order to derive the observed angle $\beta$ of the lens map in Horndeski gravity, we need to obtain the Jacobi matrix \eqref{jacobi_def} and the deflection angle $\alpha$. These should be modified by the new couplings and interactions in the gravitational sector, which were rewritten as effective stress-energy tensors related to the Einstein tensor using \eqref{field_equations_2}.

It is useful to write the Riemann tensor $R^\alpha_{\mu\beta\nu}$ in terms of the effective stress-energy tensors $T_{\mu\nu}^{(i)}$ using the field equations \eqref{field_equations_2} and its relation to its trace and trace-less parts

\begin{flalign}
\label{riemm_1}
    R_{\alpha\mu\beta\nu} &= C_{\alpha\mu\beta\nu}\\
    &+ \frac{1}{2}\left(g_{\alpha\beta} R_{\mu\nu} - g_{\alpha\nu}R_{\beta\mu} + g_{\mu\nu}R_{\alpha\beta} - g_{\mu\beta} R_{\nu\alpha} 
    \right)&&\notag\\
    &-\frac{R}{6}(g_{\alpha\beta}g_{\mu\nu} - g_{\alpha\nu}g_{\beta\mu} ).&&\notag
\end{flalign}

Using the definition of the Einstein tensor $G_{\mu\nu}$, the previous equation can be written as

\begin{flalign}
R_{\alpha\mu\beta\nu} &= C_{\alpha\mu\beta\nu}\notag\\
    &+ \frac{1}{2}\left(g_{\alpha\beta} G_{\mu\nu} - g_{\alpha\nu}G_{\beta\mu} + g_{\mu\nu}G_{\alpha\beta} - g_{\mu\beta} G_{\nu\alpha} 
    \right)&&\notag\\
    &+\frac{R}{3}(g_{\alpha\beta}g_{\mu\nu} - g_{\alpha\nu}g_{\beta\mu} ).&&\notag
\end{flalign}
In this way we can finally write \eqref{riemm_1} using the effective stress-energy tensors

\begin{flalign}
\label{riemm_2}
&R_{\alpha\mu\beta\nu} = C_{\alpha\mu\beta\nu}&&\\
    &+ \sum_2^{(5)}\frac{\left(g_{\alpha\beta} T^{(i)}_{\mu\nu} - g_{\alpha\nu}T^{(i)}_{\beta\mu} + g_{\mu\nu}T^{(i)}_{\alpha\beta} - g_{\mu\beta} T^{(i)}_{\nu\alpha} 
    \right)}{2(1+2G_4)}&&\notag\\
    &+\sum_2^{(5)}\frac{T^{(i)}}{3(1+2G_4)}(g_{\alpha\beta}g_{\mu\nu} - g_{\alpha\nu}g_{\beta\mu} )+ T^{(\text{m})}_{\mu\nu},&&\notag
\end{flalign}

where the $T^{(i)}$ are the traces of the $T^{(i)}_{\mu\nu}$. 

From the Riemann tensor \eqref{riemm_2}, we obtain a modified solution to the Jacobi matrix equation \eqref{jacobi_eq}, with the new terms involving the scalar field. We thus define the solution to this modified GDE, with Riemann tensor given by \eqref{riemm_2}

\begin{equation}
\label{GDE_eff}
    \nabla_k\nabla_k D^{A(\text{eff})}_B(\phi, X) = R^B_{\alpha \beta C} k^\alpha k^\beta D^C_A,
\end{equation}
as the effective Jacobi matrix $ D^{A(\text{eff})}_B(\phi,X)$. This Jacobi matrix therefore naturally defines the maps between lens, observer and source, as well as the angular diameter distance $d_A(z,\phi,X)$ as a function of redshift $z$ and the new kinetic and scalar couplings, for the Horndeski theories \eqref{Horndeski_luminal}.

\subsection{Distances and caustics}
Through the solution of equation \eqref{GDE_eff}, one obtains the angular diameter distances for the space-time given by the solution of the field equations \eqref{field_equations_2}. As in GR, one can define the luminosity distance \cite{Schneider} at the observer as

\begin{equation}
    \label{angular_Horn}
    d_L(z,\phi,X) = \sqrt{\det(D^{A(\text{eff})}_B(\phi, X))},
\end{equation}
which is equivalent to the definition derived from the comoving distance $\eta(z)$ for spherically symmetric metrics \cite{Perlick} 

\begin{equation}
    \label{luminosity_standard}
    d_L = (1+z)\chi(z) = (1+z)^2d_A,
\end{equation}
$\chi$ the comoving distance of the spacetime, reparametrized by $z$ and $d_A$ the angular diameter distance. This relation is commonly known as the Etherington Reciprocity relation, and its derivation can be found in, e.g. \cite{Schneider}. One must note that in Horndeski theories this does not change, as the photon number remains conserved and the geodesics are uniquely defined.

From \eqref{luminosity_standard}, one can see that, when the determinant of $D_A^B$ vanishes, distances become singular. Points $O$ and $S$ in the manifold joined by the distance $\sqrt{\text{det}D_A^B}$ and where the map $D_A^B$ vanishes non trivially are called conjugate points \cite{hawking_ellis_1973}. For a given source $S$, the light rays defined as in the previous section and mapped to the observer $O$ for which the distance is given by $d_L(z)$ may have conjugate points in its path to the observer. The set of all points conjugate to $S$ is called the caustic \cite{Schneider}.

In particular, we can write equation \eqref{jacobi_eq} as a matrix equation

\begin{equation}
\label{Jacobi_compact}
    \Ddot{\mathcal{D}} = \mathcal{RD},
\end{equation}
where
\begin{equation}
    \mathcal{R} = -\frac{1}{2}\begin{bmatrix}
    R_{\alpha\beta} k^\alpha k^\beta & 0 \\
    0 & R_{\alpha\beta} k^\alpha k^\beta 
\end{bmatrix}
+ \begin{bmatrix}
    -Re(\psi) & Im(\psi) \\
    Im(\psi) & Re(\psi) 
\end{bmatrix},
\end{equation}
and $\psi$ is defined as

\begin{equation}
    \psi \equiv -\frac{1}{2}C^\alpha_{\beta\gamma\delta}(E_\alpha^1 -iE_\alpha^2)k^\beta k^\gamma (E_\alpha^1 -iE_\alpha^2). \notag
\end{equation}

For the Horndeski terms \eqref{riemm_2}, we can expand this as to make explicit the Modified Gravity terms. Equation \eqref{Jacobi_compact} then becomes

\begin{align}
    \mathcal{R} = -&\frac{1}{2}\begin{bmatrix}
    T^{(\text{m})}_{\alpha\beta} k^\alpha k^\beta & 0 \\
    0 & T^{(\text{m})}_{\alpha\beta} k^\alpha k^\beta
\end{bmatrix} \notag\\
-&\frac{\left(G_{2X}+\Box\phi G_{3X} \right)}{4(1+2G_4)}\begin{bmatrix}
    \left(\nabla_\alpha\phi\nabla_\beta\phi\right) k^\alpha k^\beta & 0 \\
    0 & \left(\nabla_\alpha\phi\nabla_\beta\phi\right) k^\alpha k^\beta
    \end{bmatrix}\notag\\
+&\frac{2\left(G_{3\phi}-G_{4\phi\phi} \right)}{4(1+2G_4)}\begin{bmatrix}
    \left(\nabla_\alpha\phi\nabla_\beta\phi\right) k^\alpha k^\beta & 0 \\
    0 & \left(\nabla_\alpha\phi\nabla_\beta\phi\right) k^\alpha k^\beta
    \end{bmatrix}\notag\\
    -&\frac{2G_{4\phi}}{4(1+2G_4)}\begin{bmatrix}
    \left(\nabla_\alpha\nabla_\beta\phi\right) k^\alpha k^\beta & 0 \\
    0 & \left(\nabla_\alpha\nabla_\beta\phi\right) k^\alpha k^\beta
    \end{bmatrix}\notag\\
    +&\frac{2G_{3X}}{4(1+2G_4)}\begin{bmatrix}
    \left(\nabla_{(\alpha}\phi\nabla_{\beta)}X\right) k^\alpha k^\beta & 0 \\
    0 & \left(\nabla_{(\alpha}\phi\nabla_{\beta)}X\right) k^\alpha k^\beta
    \end{bmatrix}\notag\\
+ &\begin{bmatrix}
    -Re(\psi) & Im(\psi) \\
    Im(\psi) & Re(\psi) 
\end{bmatrix}.
\label{modified_matrix}
\end{align}

Here we used \eqref{field_equations_2} to rewrite the Riemann tensor in terms of the Honrdeski functions. It is not in general that a theory equations of motion can be written in the same way as in \eqref{field_equations_2}, so this method is not at all general for a given Scalar-Tensor theory. However, as long as there is a frame where the equations of motion can be separated into the form $\text{Curvature}=\text{Matter}+\text{Field interactions}$, the method is applicable. Any theory where the Einstein frame form of the lagrangian is conformally related to the non-minimally coupled form has a Jacobi matrix that can be decomposed in a way similar to \eqref{modified_matrix}, since all of the terms in the equation contractions of the Weyl tensor and null geodesics, which are both conformally invariant.

In the next subsection, we discuss how the new terms coming from the Horndeski modifications are related to the optical scalars and the focusing and distortion of light beams.

\subsection{Optical scalars and multiple imaging}
\label{multiple_lens_horn}
To uniquely solve the Jacobi Equation, one needs two initial conditions, for the value of $\mathcal{D}$ and $\Dot{\mathcal{D}}$ at the source or observer. Conventionally, one imposes the conditions at the source \cite{Perlick}, so that we understand the evolution of the quantities as a light ray past-oriented and starting at the observer, therefore inside the light cone of the observer.
In this way, we impose the conditions at the observer, which we'll call from here on the vertex, and assume that the affine parameter is $s=0$ at $\mathcal{O}$

\begin{equation}
\label{initial_conditions}
\mathcal{D}(0) = 0, \qquad \Dot{\mathcal{D}}(0) = 1.
\end{equation}

From the Jacobi matrix relation \eqref{jacobi_eq} one can define the optical scalars starting with \cite{Perlick}

\begin{equation}
    \Dot{\mathcal{D}} = \mathcal{S}\mathcal{D},
\end{equation}
where the matrix $\mathcal{S}$ is given by

\begin{equation}
\mathcal{S}=
    \begin{bmatrix}
    \theta+\sigma_1 & \sigma_2\\
    \sigma_2 & \theta - \sigma_1
    \end{bmatrix},
\end{equation}
and is called the Deformation Matrix.
 The kinematic quantities are given by $\theta$, defined as the expansion of the light bundle and $\sigma = \sigma_1 + i\sigma_2$ its shear. The geometrical interpretation of these quantities is that the expansion measures the stretching of the bundle, whereas the shear measures its distortion in the eigendirections $E_i$ of the Sachs basis \cite{Schneider}.
These quantities can be equivalently defined, such that their geometrical interpretation is more manifest, as

\begin{equation}
    \theta = \frac{1}{2}k^\alpha_{;\alpha}, \qquad \sigma = \frac{1}{2}k_{\alpha;\beta}(E_A^\alpha + iE_B^\alpha)(E_A^\beta + iE_B^\beta) 
\end{equation}
From the Geodesic Deviation Equation \eqref{GDE_eff}, and the definition of the Deformation Matrix, one obtains the Sachs Equations in Horndeski Gravity:

\begin{alignat}{3}
\label{Sachs_eq}
    &\Dot{\theta}=& -&\theta^2 -\sigma^2 -\frac{1}{2}T^{(\text{m})}_{\alpha\beta}k^\alpha k^\beta-&\\
    & & &\frac{B_1}{2}(\nabla_k\phi)^2 - \frac{B_2}{2}\frac{D^2\phi}{ds^2}-& \notag\\
    & & &\frac{B_3}{2}(2\nabla_k\phi\nabla_kX),\notag\\
    &\Dot{\sigma} =& -&2\theta\sigma-\frac{1}{2}\psi,&
\end{alignat}
where the $B_i$ are given by

\begin{align}
\label{coeff_focus}
    &B_1 = \frac{(G_{2X}+\Box\phi G_{3X}-2G_{3\phi}+2G_{4\phi\phi})}{2(1+2G_4)}, \\
    &B_2 = \frac{G_{4\phi}}{(1+2G_4)},\qquad B_3 = \frac{-G_{3X}}{(1+2G_4)}. \notag
\end{align}
Equation \eqref{Sachs_eq} is of notice, as it shows that Modified Gravity does not affect the evolution of the shear, since the Weyl tensor contractions are not modified. Therefore images are stretched in the same way as in General Relativity. One should also note that this is not frame dependent, as the Weyl tensor is preserved under conformal transformations to the Jordan frame.

Equation \eqref{Sachs_eq}, however, is modified by the extra terms arising from the effective Stress-Energy tensors. One can impose stability and energy conditions on the Horndeski functions as a restriction on the effect on the expansion and the distortion of the light beams. A discussion on Energy Conditions on Modified Gravity using the effective stress energy tensor treatment similar to the one used in this paper can be found in \cite{Capozziello_2014,Capozziello_2015}.

Here we prove a first theorem on the properties of multiple lensing and the effect of the modification of gravity. We follow closely the arguments presented in \cite{padmanabhan_88} and \cite{Subramanian_86}, and use the results presented in \cite{hawking_ellis_1973}, section 4.4 on conjugate points.

\begin{theorem}

\label{theo_1}
Suppose that the matter stress-energy tensor satisfies the null energy condition, and that the Horndeski functions satisfy $B_1(\nabla\phi)^2 +B_2(\frac{D^2\phi}{ds^2})+2B_3(\nabla_k\phi\nabla_kX) \geq 0$ on the light bundle generated by $k^\mu$. Then the following statements are true:
\begin{itemize}
    \item The lens produces multiple images.
    \item If the scalar field is smooth and bounded at the lens, then the number of images is the same as in General Relativity.
\end{itemize}
\end{theorem}

\begin{proof}
First we note that there is no loss in generality in redefining $(1+2G_4)$ as $G_\text{eff}$, and ask that it is positive.
Thus, assuming that the Horndeski functions satisfy the mentioned conditions, the right hand side of \eqref{Sachs_eq} is strictly negative.
Note, from the definition of $\theta$ and the luminosity distance \eqref{luminosity_standard} that 

$$ \theta = \frac{\dot{d_L}}{d_L}, $$
and thus that if $\theta\rightarrow \infty$ then $d_L\rightarrow 0$.

From the negativity of $\Dot{\theta}$, and the initial conditions $\dot{d_L}=1$, there must be a point where $\theta < 0$. Then there is a conjugate point to the observer, applying the mean value theorem for integrals, Proposition 4.4.1 of \cite{hawking_ellis_1973}.

The existence of a conjugate point to the observer guarantees that there are multiple images from the effect of the lens, following the main theorem of \cite{padmanabhan_88}. This proves the first item.

From the assumption that the scalar field is bounded at the lens,the total amount of energy density of the lens must be bounded, as the effect of the scalar field is limited. Then the lensing angle is bounded \cite{Schneider}. Therefore, as argued in \cite{Subramanian_86}, there is not only multiple imaging, but the number of images is odd exactly as in GR, as per the result of Burke's theorem.
\end{proof}

The result of the previous theorem shows that, for a space-time under the same energy conditions as in General Relativity, we don't expect different behavior in Modified Gravity as long as the coefficients \eqref{coeff_focus} obey certain inequalities. We proceed to apply the theorem to some of the theories described in \cite{Kase_2019}, and note that in the notation of our paper, $\mathcal{L}^{(3)} = -\Box\phi G_3(\phi,X)$ and $\mathcal{L}^{(4)} = (G_4(\phi,X)-1/2)R$, such that we take these factors into account in the Lagrangians for the models described below.

\subsubsection{f(R) and Brans Dicke theories}
$f(R)$ theories can be mapped, both in the metric and Palatini formalism, to Brans-Dicke theories with Brans Dicke parameter $\omega\geq0$.
For this kind of theory, one has the Horndeski functions

\begin{align}
    &G_2 = \omega\frac{X}{\phi}, \quad G_3 = 0, \quad G_4 = \frac{\phi}{2}-\frac{1}{2},\\
    \implies &B_1 = \frac{\omega}{2\phi^2}, \quad B_2 = \frac{1}{2\phi}, \quad B_3 =0,
\end{align}
such that, in order to satisfy the theorem, one needs the condition
\begin{equation}
    \frac{\omega}{\phi^2}(\nabla\phi)^2 \geq -\frac{1}{\phi}\frac{D^2\phi}{ds^2}.
\end{equation}
One can take $\phi$ as positive, which guarantees stability of solutions and nondegeneracy of the equations of motion. The previous equation is then simplified to

\begin{equation}
    \omega\frac{(\nabla\phi)^2}{\phi}\geq -\frac{D^2\phi}{ds^2}.
\end{equation}

For metric $f(R)$ theories, the BD parameter is $\omega = 0$, and the condition is satisfied if the second derivative of the scalar field is nonnegative on the geodesic. For Palatini $f(R)$, which corresponds to $\omega = -3/2$ with a potential term, one needs that $(\nabla\phi)^2/\phi\leq 2/3\frac{D^2\phi}{ds^2}$. For arbitrary Brans Dicke theories as long as $\omega \gg 1$, one can guarantee that the condition is satisfied; this limit is usually regarded as the GR limit of the theory. For cosmological models, which are our main interest, this range of parameters is currently allowed by observations \cite{Clifton_2012}.

\subsubsection{Galileon Ghost Condensate}
The Horndeski functions for the Galileon Ghost Condensate model, which allows for phantom crossing in dark energy's equation of state through a nonlinear kinetic term \cite{Peirone_2019}, are given by

\begin{align}
    &G_2 = a_1X+a_2X^2, \quad G_3 = -3a_3X, \quad G_4 = \frac{M_{\text{Pl}}^2-1}{2},\\
    \implies &B_1 = \frac{(a_2X+2a_1+3a_3\Box\phi)}{M_{\text{Pl}}^2}, \quad B_2 = 0, \quad B_3 =\frac{a_3}{M_{\text{Pl}}^2}.
\end{align}
This theory has a nontrivial coupling to the $G_3$ part of the action, which is related to the cubic interaction. 
The condition for this theory to satisfy the theorem is then

\begin{equation}
    (a_2X+2a_1+3a_3\Box\phi)(\nabla\phi)^2+a_3(\nabla_k\phi\nabla_kX)\geq 0,
\end{equation}

Assuming a flat RW metric, the above equation can be rewritten as 

\begin{equation}
    a_2\dot{\phi}^4+2a_1\dot{\phi}^2+3a_3\Ddot{\phi}\dot{\phi}^2+a_3\dot{\phi}^3\geq 0,
\end{equation}

which is directly related to the observational parameters $x_i$ related to the dark energy density defined in 
\cite{Peirone_2019,Peirone_2_2019}. 
With the stability assumptions $\dot{\phi},\Ddot{\phi}> 0$ and using the observational constraints from \cite{Peirone_2019,Peirone_2_2019}, we have $a_1, a_2, a_3 \geq 0$, and the inequality is identically satisfied. Thus, for cosmological settings, the current observational constraints reproduce the observed gravitational lensing behavior, as predicted by GR.


\subsubsection{Unified k-essence}

Unified k-essence was first proposed in \cite{Scherrer_2004} as a scalar field model unifying dark energy and dark matter through a single scalar field with quadratic kinetic term, with Horndeski functions given by \cite{Kase_2019}

\begin{align}
    &G_2 = -b_0+b_2(X-X_0)^2, \quad G_3 = 0, \quad G_4 = \frac{M_{\text{Pl}}^2-1}{2},\\
    \implies &B_1 = \frac{2b_2(X-X_0)}{M_{\text{Pl}}^2}, \quad B_2 = B_3 =0.
\end{align}

The $X_0$ is a positive constant characteristic kinetic scale, the extremum of the function $G_2$ \cite{Scherrer_2004}.
The requirement on the functions to satisfy the theorem is then

\begin{equation}
\label{unified_kess}
    \frac{2b_2(X-X_0)}{M_{\text{Pl}}^2}(\nabla\phi)^2 \geq 0.
\end{equation}

In order for this theory to reproduce the matter epochs in a cosmological setting, one requires that $X-X_0 \approx X_0(1+\epsilon(t))>0$ \cite{Kase_2013,Scherrer_2004}, so the constraint \eqref{unified_kess} is satisfied. This is in agreement with Bekenstein and Sanders' result that gravitational lensing in Scalar-Tensor theories which try to account for the dark matter effect cannot significantly modify the results derived by General Relativity \cite{Bekenstein_1994}.

\subsubsection{Generalized Brans-Dicke}
In this model with non-trivial cubic and nonminimal coupling, introduced in \cite{De_Felice_2011}, the cosmological and stable solutions possess Horndeski functions \cite{Kase_2019,De_Felice_2011}

\begin{align}
    &G_2 = \omega\left(\frac{\phi}{M_{\text{Pl}}}\right)^{1-n}X, \quad G_3 = -\frac{\lambda}{\mu^3}\left(\frac{\phi}{M_{\text{Pl}}}\right)^{-n}X,\notag\\
    &G_4 = \frac{M_{\text{Pl}}^2}{2}\left(\frac{\phi}{M_{\text{Pl}}}\right)^{3-n} - \frac{1}{2},\\
    \implies &B_1 = \left(\frac{\phi}{M_\text{Pl}}\right)^{-2}\Bigg[\frac{\omega}{M_\text{Pl}}+\frac{\lambda\Box\phi}{\mu^3 M_\text{Pl}} \left(\frac{\phi}{M_\text{Pl}}\right)^{-1}\notag\\
    &-4n\frac{\phi}{\mu^3}\left(\frac{\phi}{M_\text{Pl}}\right)^{-1}+(3-n)(2-n)\phi^{-2}\left(\frac{\phi}{M_\text{Pl}}\right)^{-1}\Bigg]\notag \\
    &B_2 =\frac{(3-n)}{2}\phi^{-1}, B_3 = \frac{\lambda}{\mu^3M_{\text{Pl}}^2}\left(\frac{\phi}{M_{\text{Pl}}}\right)^{-3},
\end{align}

with the parameter $n$ satisfying $2\leq n \leq 3$ and the couplings satisfying $\omega < 0$, $\lambda >0$  and $\mu>0$ \cite{De_Felice_2011}. For this theory, the condition is not necessarily satisfied, as its validity is highly dependent on the parameter values. In the case $n=3$, the condition becomes

\begin{align}
    &\left(\frac{\phi}{M_\text{Pl}}\right)^{-2}\left[\frac{\omega}{M_\text{Pl}}+\frac{\lambda\Box\phi}{\mu^3 M_\text{Pl}} \left(\frac{\phi}{M_\text{Pl}}\right)^{-1}-12\frac{\phi}{\mu^3}\left(\frac{\phi}{M_\text{Pl}}\right)^{-1}\right](\nabla\phi)^2+\notag\\
    &\frac{\lambda}{\mu^3M_{\text{Pl}}^2}\left(\frac{\phi}{M_{\text{Pl}}}\right)^{-3}(\nabla_k\phi\nabla_kX)\geq 0,
\end{align}
which is more tractable, although still dependent on the theory's parameter space. In particular, since the parameters are also dependent on the late time behavior of cosmological solutions, one could in principle test the behavior of the theory through numerical solutions of the field equations with given cosmological parameters, as done in \cite{De_Felice_2011}.

\subsection{Focusing and magnification}
In General Relativity, the focusing theorem guarantees that for the most general spacetimes satisfying the weak energy condition, the Gravitational potential has a focusing effect, that is, null rays forming an infinitesimal bundle converge when passing through a gravitational lens \cite{Perlick}. Equivalently, the cross section with angular size $\delta\theta$ of the image generated by a source $S$ gets smaller as the light passes through the gravitational lens.
Since the angular size of the cross section is related to the luminosity distance $d_L(z)$ through the Etherington relation for the angular diameter distance $\delta\theta = \delta l/d_A(z) = (1+z)^2 \delta l/d_L(z)$, where $dl$ is the object true observed size, the evolution of the luminosity distance modifies the cross section.

One can define, for a light bundle with cross-section $\delta\theta$ at the source, the magnification factor $\mu$ at the observer, which is given by \cite{Perlick}

\begin{equation}
    \label{magnification_fac}
    \mu = \frac{\delta\theta}{d_L^2} = \frac{s^2}{d_L(s)^2},
\end{equation}
where $s$ is the affine parameter of the bundle, and we have used that $d_L(s) \approx s$ near geodesics vertexes. The infinitesimal area of the bundle is $\delta \theta \approx s^2$, as one can check from the definition of the geodesic bundle in \eqref{bundle}.

From the Sachs equations \eqref{Sachs_eq} and the definition of the optical scalars, one can write the Focusing Equation

\begin{equation}
\label{focusing_eq}
\Ddot{d_L} = -\left(|\sigma|^2 + \frac{1}{2}R_{\alpha\beta}k^\alpha k^\beta\right)d_L,
\end{equation}

and the Focusing Theorem is the statement that $d_L(s) \leq s$. It follows from the integration of the previous equation on both sides and the initial conditions defined in \eqref{initial_conditions}. In General Relativity, one just needs the weak energy condition for the right hand side of \eqref{focusing_eq} to be strictly nonpositive. The immediate consequence is that

$$\mu =  \frac{s^2}{d_L^2}\geq 1, $$
such that light-beams are focused when passing through the lens, or that areas are magnified.

For Horndeski theories, one obtains the Modified Focusing Equation
\begin{align}
\label{focus_eq_mod}
    \Ddot{d_L} =& \notag\\
    -\bigg(&|\sigma|^2 + \frac{1}{2}T^{(m)}_{\alpha\beta}k^\alpha k^\beta+\frac{B_1}{2}(\nabla_k\phi)^2 +\notag\\ &\frac{B_2}{2}\frac{D^2\phi}{ds^2}+\frac{B_3}{2}(2\nabla_k\phi\nabla_kX) \bigg)d_L.
\end{align}

Under the conditions of the previous theorem, one can see that the Focusing theorem is easily satisfied, since the right hand side of \eqref{focusing_eq} is strictly non-positive and integrating both sides twice on the affine parameter $s$ of the geodesic.

The previous condition, however, is not necessary but sufficient. A weaker condition on the Horndeski functions is the one on the following theorem

\begin{theorem}
\label{theo_2}
Suppose that the matter stress-energy tensor satisfies the null energy condition, the initial conditions \eqref{initial_conditions} are valid, and that the Horndeski functions satisfy $\int_0^s[B_1(\nabla\phi)^2 +B_2(\frac{D^2\phi}{ds^2})+2B_3(\nabla_k\phi\nabla_kX)]ds \geq 0$ on the light bundle generated by $k^\mu$. Then any image that passes through that lens is magnified, that is

$$\mu \geq 1. $$
\end{theorem}

\begin{proof}
Integrating both sides of \eqref{focus_eq_mod}, one gets

\begin{flalign}
    \int_0^s  \Ddot{d_L}(s)ds \leq  -\int_0^s \Bigg[\bigg(|\sigma|^2 +\frac{1}{2}T^{(m)}_{\alpha\beta}k^\alpha k^\beta&\notag\\
    +\frac{B_1}{2}(\nabla_k\phi)^2 + \frac{B_2}{2}\frac{D^2\phi}{ds^2}+\frac{B_3}{2}(2\nabla_k\phi\nabla_kX) \bigg)d_L(s)\Bigg]&\leq 0\notag\\
    \Rightarrow \Dot{d_L}(s)-1& \leq 0\notag\\
    \Rightarrow d_L(s)& \leq s
\end{flalign}
\end{proof}
The condition for this theorem is sometimes called the averaged energy condition \cite{Fewster07}, applied to the effective stress-energy tensor.
The conditions of \ref{theo_2} are much weaker than the ones in \ref{theo_1}, as one doesn't need that the functions in the right hand side of  \eqref{focus_eq_mod} be strictly non-negative, rather that their integral be strictly non-negative. Trivially, if a class of theories satisfies the conditions of \ref{theo_1}, it also satisfies the conditions of \ref{theo_2}.

For the theories discussed in \ref{multiple_lens_horn}, where the validity of \ref{theo_1}, nothing changes in relation to \ref{theo_2}. The interesting cases are the ones where the dependence on parameters avoided the validity of the theorem. Now that the condition is over the average of the scalar field dynamics on the null geodesics, as long as the dynamics preserves the left hand side of \eqref{focus_eq_mod}, one does not need to impose that the functions $B_i$ don't change sign.

For the Generalized Brans Dicke and Ghost Condensate theories discussed in the previous subsection, numerical analysis of the cosmological dynamics could give a range of parameters where the theorems are valid. One could also use the observation of lenses as a test to the parameter range of the theories. Once one is able to use cluster and galaxy lensing statistics to constrain the magnification effect, this could put a constraint on the allowed parameters of theories that violate the average conditions, although this would need numerical evaluation of the focus and magnification equations.

In the case of Unified k-essence, the fact that the theory does not predict a deviation from the magnification derived by GR is in accordance with the results obtained in \cite{Bekenstein_1994} in relation to Scalar field dark matter models. Although the case was made not for an accelerating cosmological model, it supports the understanding that lensing is not quantitatively modified by the inclusion of minimally coupled scalar fields. The definitive results would need a quantitative result of equation \eqref{focus_eq_mod} for cosmological models, which we leave for future work. 

\section{Discussion and remarks}
\label{conclusions}
In this paper we have developed general mathematical results one can use to test and understand gravitational lensing in theories of the luminal Horndeski type. Theorems \ref{theo_1} and \ref{theo_2} impose sufficient conditions these theories must satisfy such that the effect of strong gravitational lensing is the same as in General Relativity. We examined these conditions and obtained inequalities the theory parameters need to satisfy, sometimes trivially in physical cases, such as metric $f(R)$\cite{Sotiriou_2010} and Unified $k$-essence \cite{Scherrer_2004}, which shows that some classes of theories should not modify the qualitative behavior of lensing at all.

From this formalism, one could in principle derive numerical results, using equations \eqref{lensing_horndeski} and \eqref{focus_eq_mod}, to further constrain the theory parameter space where the gravitational lensing behavior doesn't deviate from General Relativity. Together with the calculation of the bending angle, found for instance in \cite{Bekenstein_1994}, one could derive statistics from multiple strong lensing systems and constraint the deviation from General Relativity without phenomenological or effective approaches. The observation of gravitational lensing in this regime is then able to directly constrain the parameter space of scalar tensor theories.

The formalism in section \ref{lensing_horndeski} is general and applies not only to strong gravitational lensing, but to any lensing regime. Another possible useful application of this formalism is the study of gravitational weak lensing in cosmological settings, which is of particular interest in the EFTDE \cite{Frusciante_2020}, where one can relate the Horndeski functions in \ref{Horndeski_EFE} to observable cosmological parameters obtained from perturbation theory, differentiating the effects arising from the Modified Gravity models from pure shear and convergence effects. The formulation of the Horndeski interactions as effective stress-energy tensors allows the test of phenomenological descriptions of dark energy with little modification to the equations. The effect of a cosmological constant on strong gravitational lensing can then be tested using different approaches from the one found in e.g. \cite{Bessa_22}.

We find that the gravitational lensing effect in Modified Gravity is qualitatively identical to the one in General Relativity for popular models of Modified Gravity such as metric $f(R)$ and Unified $k$-essence. For other theories, we've shown that requiring the validity of the theorems constrains their parameter space through the Horndeski functions of the theory. Precise constraints can be obtained assuming a given lens model and observations, and imposing the condition that lensing should not deviate from GR predictions. 

Using the bending angle predictions for these theories \cite{Bekenstein_1994,Keeton2005}, together with the constraints obtained in this paper, one can use strong lensing systems to test Modified Gravity models. In the next decade the amount of cluster and galaxy lensing data is expected to increase by orders of magnitude \cite{Bom_2022,Collett_2015}. This new batch of data can provide new statistics once we're able to precisely constraint the lens models in order to separate relativistic effects from the Modified Gravity ones. We leave analysis of this kind for future work.

\section*{Acknowledgments}
Pedro Bessa would like to thank FAPES and CAPES
for the PhD scholarship, as well as CBPF and Université de Genève for providing office space and computational power. He would also like to thank Marcela Campista and Alexsandre Ferreira for comments and reviews on the manuscript.

\bibliography{main} 
\bibliographystyle{unsrt}

\nocite{*}

\end{document}